\documentclass[aps,
twocolumn, showpacs]{revtex4}
\usepackage{natbib}
\usepackage[printonlyused]{acronym}
\usepackage{graphicx}
\usepackage{epstopdf}

\usepackage{ifthen}
\usepackage{color}
\newcommand{\comment}[2][keb]
	{
	\marginpar{
		\flushleft\scriptsize\setlength{\baselineskip}{7pt}
		\ifthenelse{\equal{#1}{keb}}
			{\color{green}}
			{}
		\ifthenelse{\equal{#1}{mcw}}
			{\color{red}}
			{}
		\ifthenelse{\equal{#1}{jrt}}
			{\color{blue}}
			{}
		{\MakeUppercase{#1}: #2}
		}
	}

\begin{document}

\title{Visualizing Changes in Student Responses Using Consistency Plots}
\date{\today}
\author{Katrina E. Black}
\affiliation{Department of Physics and Astronomy, University of Maine, Orono, ME 04469}
\author{Michael C. Wittmann}
\email{wittmann@umit.maine.edu}
\affiliation{Department of Physics and Astronomy, University of Maine, Orono, ME 04469}
\affiliation{Center for Science and Mathematics Education Research}
\affiliation{College of Education and Human Development}

\begin{abstract}
Traditional methods of reporting changes in student responses have focused on class-wide averages. Such models hide information about the switches in responses by individual students over the course of a semester. We extend unpublished work by Steven Kanim on ``escalator diagrams" which show changes in student responses from correct to incorrect (and vice versa) while representing pre- and post-instruction results on questions. Our extension consists of ``consistency plots" in which we represent three forms of data: method of solution and correctness of solution both before and after instruction. Our data are from an intermediate mechanics class, and come from (nearly) identical midterm and final examination questions. 
\end{abstract}

\pacs{01.40.Fk, 01.40.gf}

\maketitle
\acresetall

\section{Introduction}

At the University of Maine, we are investigating what conceptual, procedural, and epistemological tools students use when solving first-order, separable differential equations in the context of intermediate mechanics \cite{Ambrose2004,Wittmann2007b}. As part of this study, we are interested in how student responses to identical questions change over time. 

In physics education research (PER), we often compare student results on identical questions. So, for example, we have pre-instruction and post-instruction use of surveys like the Force Concept Inventory (FCI) \cite{Hestenes1992} and Force and Motion Conceptual Evaluation (FMCE) \cite{Thornton1998}. Questions from an ungraded quiz might come back in only slightly altered form on later examinations. Many instructors use questions twice: one question from the midterms might show up on the final examination. Furthermore, we often claim that questions asked before instruction as ``baseline" data for studying the level of student understanding are nearly identical to later, post-instruction examination questions (see, for example, many of the citations in ref. \cite{McDermott1999}). We often compare students' responses on each question to then make claims about the effectiveness of a teaching intervention in the process.

Changes in student performance (as from pre- to post-instruction tests) have typically been reported on a class average basis (see, again, many of the citations in ref. \cite{McDermott1999}). This method of reporting has been very useful in roughly gauging the overall knowledge state of the class when assessing the effectiveness of a different kinds of curricula \cite{Hake1998,Redish1999}. However, comparing class averages can only shed light on individual student response patterns when the individual response changes like the group response. When class response patterns remain static, we know next to nothing about individual student responses. To describe the space between these two extremes, Bao \cite{Bao2006a} discusses the differences between reporting the class average normalized gain and the average of individual studentsÕ normalized gains and suggests methods by which the differences in these two measures can yield information about the students under discussion. Further issues arise, for example static group data in which individuals give varying and internally inconsistent responses. Such differences can have meaning \cite{Bao2006}.

\begin{figure}
\begin{center}
\includegraphics[width=8cm]{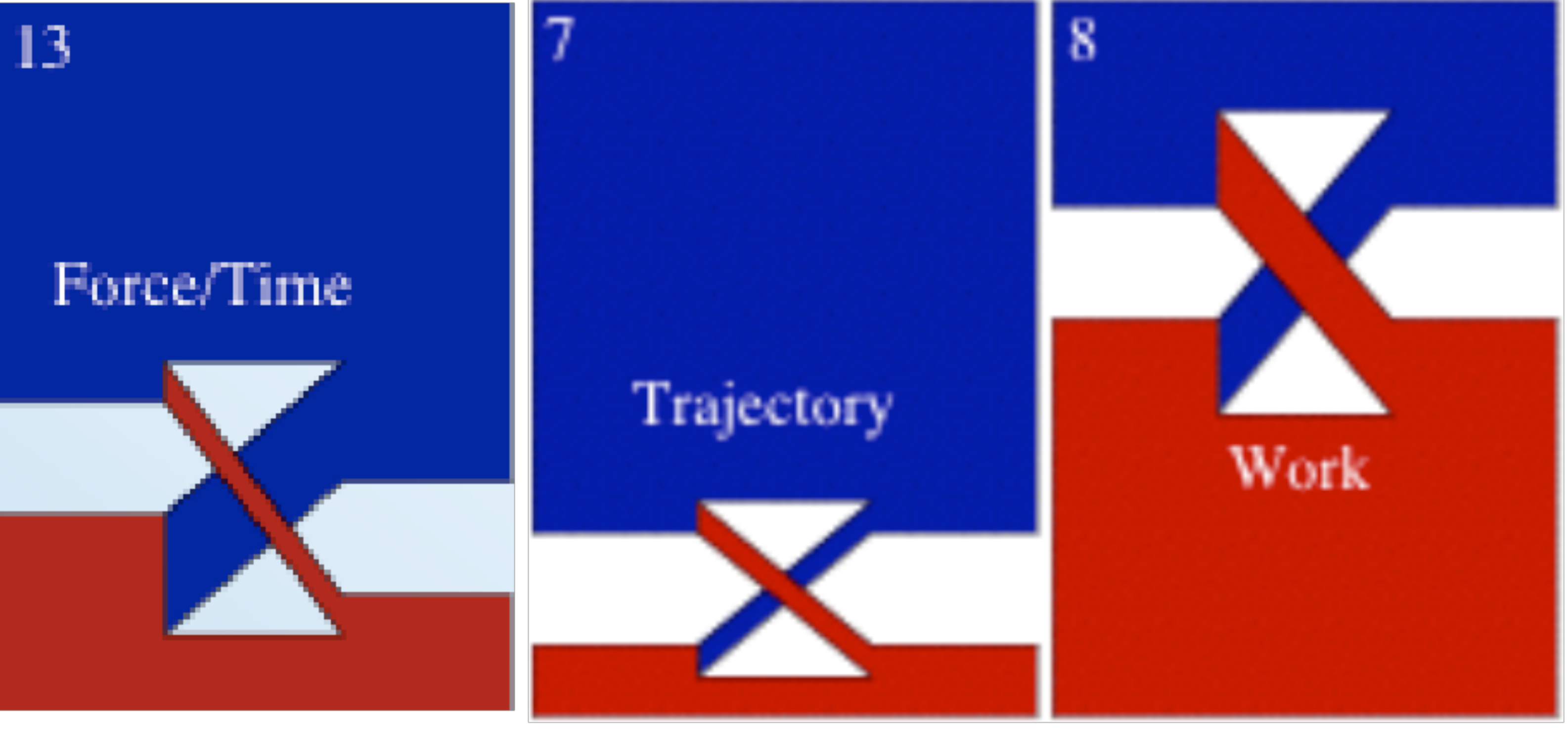}
\caption{Caption for Kanim escalator graph.}
\label{Escalator}
\end{center}
\end{figure}

Others have attempted to address this inherent difficulty in reporting changes in student response. The goal, as introduced by Kanim (unpublished), was to indicate class-wide shifts from more correct to more incorrect responses, without assuming that no students moved from incorrect to correct. Kanim developed the visually compelling escalator diagram (see figure \ref{Escalator}) that illustrates in a compact icon both how many students responded correctly or incorrectly to a particular question at two different times and how many students changed their response from correct to incorrect or vice versa \footnote{Kanim described escalator plots as part of a presentation in December, 2007, at the University of Maine, but they were developed previously.}. The plot is read from left to right. The size of the blue and red regions on the left edge indicates how many students got the question correct or incorrect during the first asking; the size of the corresponding regions on the far right represent correct and incorrect answers during the second asking. While most students maintain their correctness state, some students who were initially correct get on the red ``down escalator" and give an incorrect answer later, and some who were initially wrong go up the blue escalator. The width of the diagonal escalator lines indicates how many students changed their answer.

Figure \ref{Escalator} shows three qualitative escalator diagrams. In the first, ``Force/Time", more students change to correct than incorrect, leaving a net positive change in correctness. In the second and third, the total number of students answering correctly remains the same. This, obviously, requires that an equal number of students change their answers from correct to incorrect and incorrect to correct between the two administrations of the test, illustrated by the equal width diagonal lines connecting the blue (correct) and red (incorrect) regions. However, in the ``Trajectory" diagram, the number of students changing their answers is relatively small, whereas in the ``Work" diagram, a relatively large number of students change their answers. This information would be obscured in a traditional tabular representation of the data. 

\begin{figure*}
\begin{center}
\includegraphics[width=6in]
{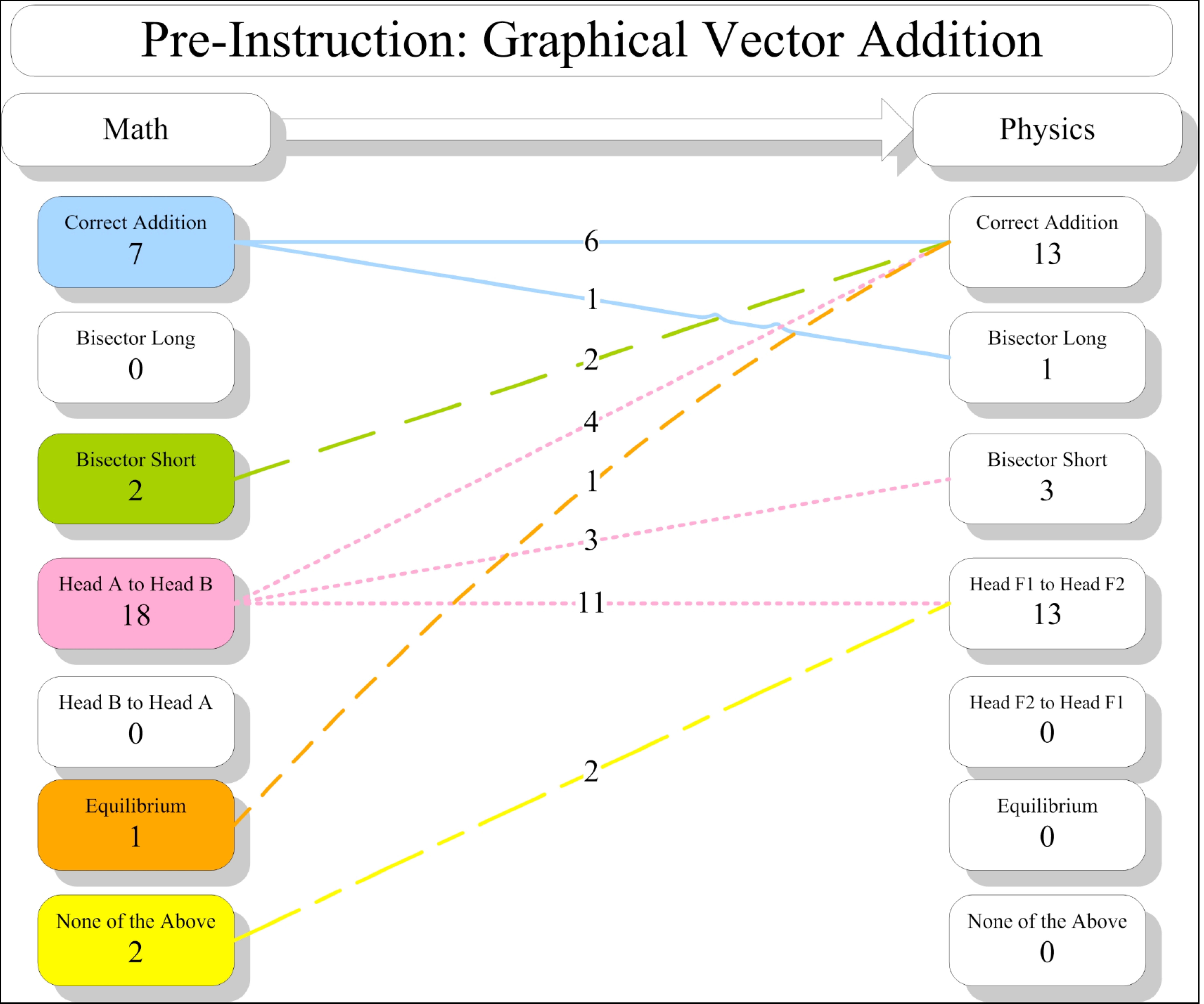}
\caption{Expanded Escalator Plot. By including all categories of partially incorrect responses (most egregious at the bottom), we gain more understanding of student responses on nearly identical questions (see ref. \cite{VanDeventer2008} for more details).}
\label{JoelEscalator}
\end{center}
\end{figure*}

In an escalator diagram, the ``red" answer can be either the most common incorrect response (leaving some ``other" as white on the diagram) or it can be all incorrect answers lumped together (and the white area remains to help visualize the situation better). Van Deventer \cite{VanDeventer2008} extended the escalator diagram representation to include many different responses to a question. Figure \ref{JoelEscalator} shows an example from ref. \cite{VanDeventer2008} on student responses to \textit{2-d} vector addition questions in a physics context that came after nearly identical questions in a non-physics ``math" context. In this paper, we take Kanim's philosophy and extend it even further to include additional information. 

Many kinds of problems in physics have more than one correct method of solution. For example, you can use both energy reasoning or force reasoning to correctly solve certain problems in mechanics. As researchers, we often wish to connect student solution methods with an analysis of the correctness of a solution: Are the students using method A doing better than those using method B?

The topic discussed in this paper is air resistance, of interest because it contains the relatively simple analysis of a first-order differential equation. We asked identical air resistance problems on a midterm and final and described student responses in two ways. In one description, we coded the work for the \textit{technique} the student used. In another description, we coded responses based on the \textit{type of mistakes} students made. To compare these two complementary descriptions of individual student method, along with how these methods changed in time, required a more visual solution than the typical table, and was too complex to make efficient use of escalator diagrams. This led to the development of consistency plots, a kind of \textit{2-d} escalator diagram. There are fully described later in the paper. 
	
In the following section, we discuss the exam question analyzed and our coding methodology, followed by a presentation of our results in a typical table format. In Section \ref{Plots}, we describe consistency plots in general and present a plot of our data. In particular, in section \ref{Elements}, we discuss specific student response patterns which are obscured by a tabular presentation of the data. We discuss some limitations of the consistency plot representation in section \ref{Difficulties}.
	
\section{\label{Question}Student responses to a physics question}

One of the first problem types in mechanics that requires the solution of a first-order differential equation is the air-resistance problem for objects in a gravitational field. We present three years of data from an examination question that appeared both on a midterm early in the semester (immediately following instruction on air resistance) and later on the final examination. In 2006, air resistance appeared in the first few weeks of the course, following a brief review of introductory mechanics. Since about half of our students take a differential equations course in the math department concurrently (the other half have taken it previously), we wanted to allow these students to encounter the technique of separation of variables in a mathematical context before doing so in a physics context.  In 2007 and 2008, air resistance was covered in the middle of the semester. 

\subsection{The exam question and solution}

\begin{figure*}
\begin{center}
\includegraphics[width=6in]{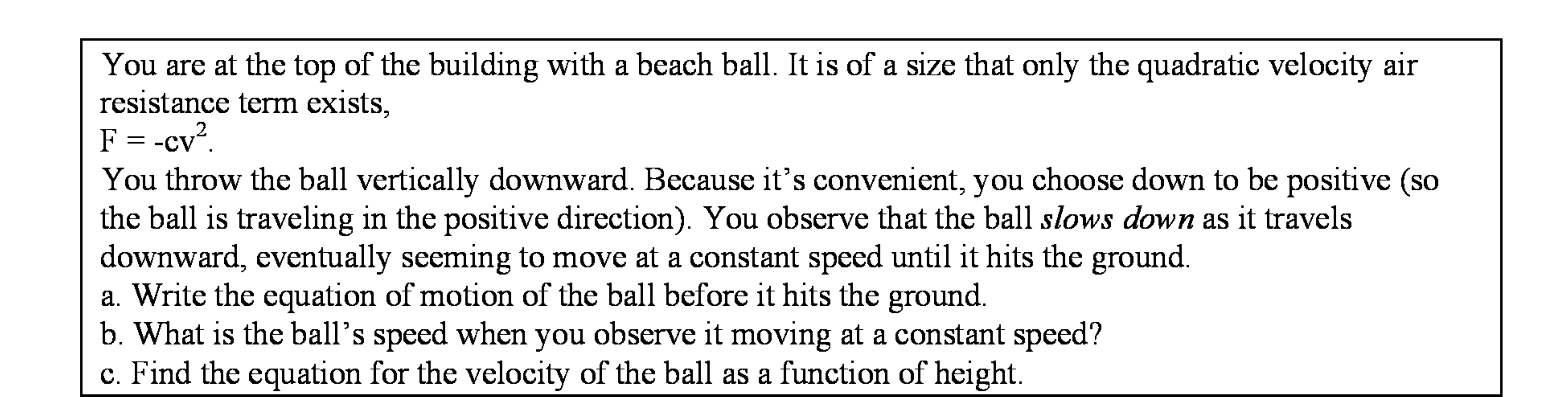}
\caption{Problem statement. Only the vertical motion part is shown.}
\label{Problem}
\end{center}
\end{figure*}

In 2006, the question (see figure \ref{Problem}) appeared in a midterm exam given in Week 4 of the course. In 2007 and 2008, it appeared as a group quiz in the middle of the semester. Students in 2007 and 2008 worked on the problem in small groups during class time, and then solved the problem again on their own in an individual, take-home component of the quiz. We present results from these individual responses rather than the videotaped group quizzes. We look in particular at studentsÕ solutions to part \textit{c} of the problem. We note that issues related to the videotaped group quiz are discussed in ref. \cite{McCann2009} and deal with issues related to coordinate systems, not discussed in detail in this paper.

\begin{figure}
\begin{center}
\includegraphics[width=4.5cm]{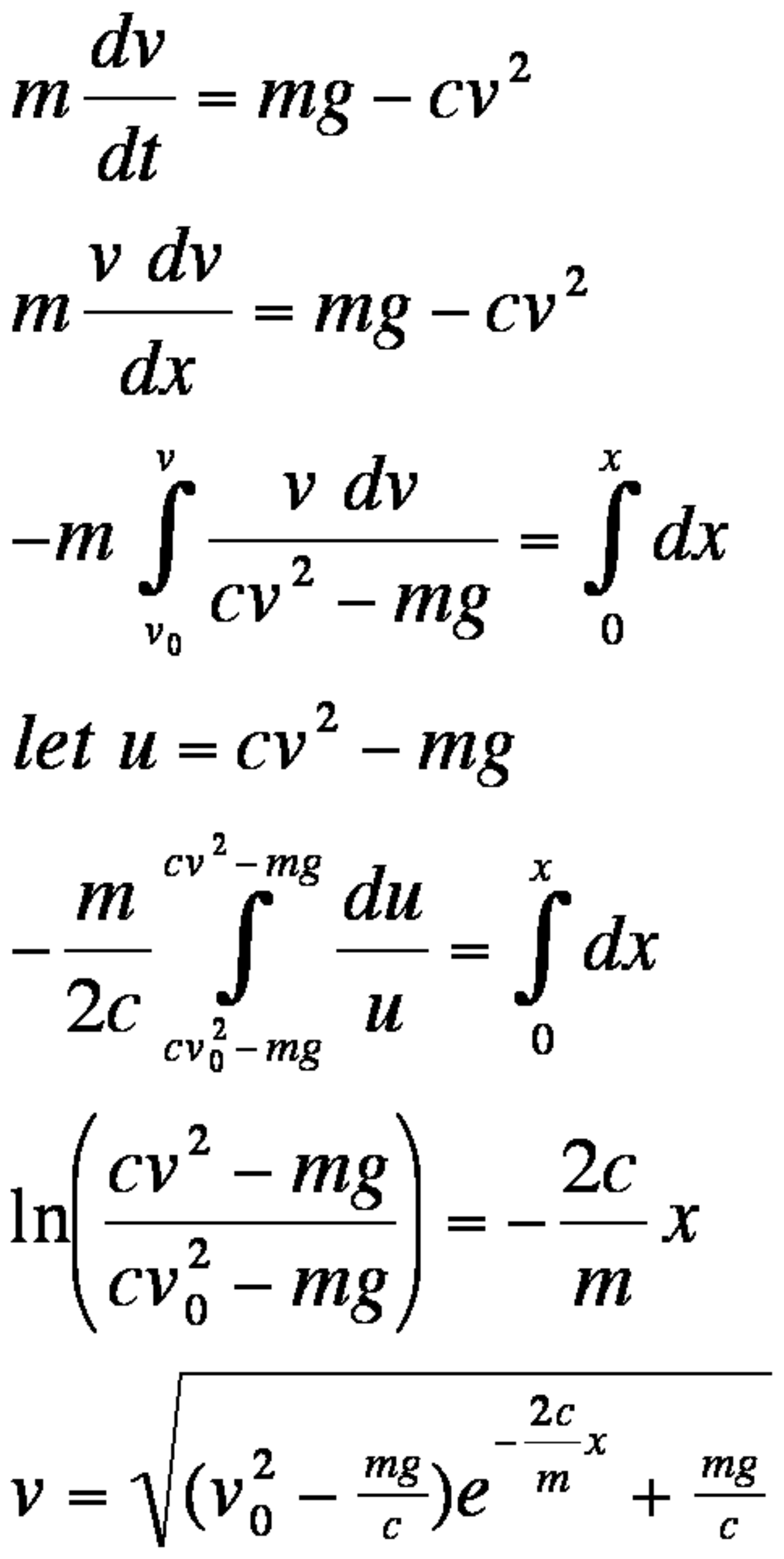}
\caption{Problem solution using limits method. Explanations have been omitted for brevity.}
\label{Derivation}
\end{center}
\end{figure}

A correct solution requires several important steps. First, students must apply the correct coordinate system, with down as the positive direction and the origin at the launch height. Second, a coordinate transformation from $a = dv/dt$ to a differential based on displacement, $a = v dv/dx = 1/2 d(v^2)/dx$, is required. In 2006, this transformation was part of the previous problem on the exam, providing a hint for students who might get stuck. Once the transformation had been applied and variables separated, the problem is simplified with a \textit{u}-substitution that allows the \textit{v}-integral to be solved with relative ease. Finally, students had to either choose the correct limits of integration (\textit{limits method}) or add an integration constant and find its value (\textit{+C method}). Both methods are correct, of course. An example solution of only the mathematical steps involved in using the limits method is shown in figure \ref{Derivation}.

In this problem, the initial velocity was not given explicitly and had to be defined by the students as $v_0$ or something similar. Also, the question asked for students to give the velocity as a function of height (as measured from the top Ð not a typical definition of height!) even though they had just found the terminal velocity. Although this might have impaired some studentsÕ reasoning, evidence from video recordings of the group quizzes (not discussed in this paper) suggests that this atypical definition was not generally problematic for the students, who seemed to use height and distance interchangeably.
	
The same question was given unchanged on the final examination in 2006 and 2008. In 2007, the question was changed slightly. Students were asked to find the velocity as a function of time rather than position, and the air resistance force was $v$- and not $v^2$-dependent. More consequentially, students were to consider up as the positive direction but to consider both upward and downward motion. This slight perturbation led to a large effect in the data, creating the necessity of sometimes excluding or recording separately the results from this year. We will indicate each time this is the case. The dependence of student reasoning on simple changes to coordinate systems is an interesting point, and consistent with other results from our studies on student use of coordinates and coordinate systems, but outside of the scope of this paper \cite{Sayre2006,Sayre2007a,Sayre2008,McCann2009}. Additionally, because we are discussing changes in individual student responses, we will consider only those students (in all years) for whom we have matched data. Of the 39 students who took both exams, six students had a completely correct solution on the midterm and seven were completely correct on the final. We note that solutions to differential equations were not a large part of the course content in the last third of the semester. We present a more detailed analysis of the results in the following section.

\subsection{\label{Tabular}Tabular presentation of student answers}

We have gathered data from three years of instruction. Solution method categories arose from the data, being those methods most commonly used. Most students used separation of variables to solve the DE. These students were placed into one of the three groups: the \textit{limits method} common to physics classes, the \textit{+C method} more common to math contexts, and a \textit{both limits and +C} method in which students applied both methods to the problem. Those students who did not use separation of variables (including two students who applied the more complex technique of variation of parameters with an integrating factor) as well as students who used separation of variables but did not use limits or an integration constant (essentially, only finding the anti-derivatives) were categorized as ``other" solutions. As might be expected, most students who did not use separation of variables were unable to progress far into the problem.

We also grouped solutions according to correctness. ``Correct" solutions were carried out mathematically correctly and related the mathematics to the physical meaning by successfully using the initial conditions in the final statement of the solution. Problems with a ``math error" could include simple algebraic mistakes or might progress to difficulties using a \textit{u}-substitution. Students who failed to consider the domain of the natural log function were also placed in this group, even though the reasoning required to determine the sign of the argument of the $ln$ function was physical \footnote{As the problem is stated, a common error is that the argument of the \textit{ln} function is negative, and therefore undefined in an intermediate step of the problem}. Those students with a ``boundary condition error" might use unphysical limits, set $v_0 = 0$ later in the problem, or leave the integration constant undefined, simply as \textit{+C}. Finally, students in the ``other"  category either did not come to a final answer or made a serious physical error while setting up the problem (such as writing terms in the equation in a way that contradicted the given coordinate system \cite{McCann2009}).

Our original goal in this analysis was to observe how students used boundary conditions to give physical meaning to mathematical statements. Therefore, we defined the ``boundary condition error" category very narrowly and defined the ``math error" and ``other" categories quite broadly. For our specific purposes, math errors of any type are less serious than boundary condition errors because the latter depend on the students' physical reasoning abilities while the former may simply be evidence of a careless math error. Different research goals (say, on the use of coordinate systems and their role in translating physical systems into mathematical statements \cite{Sayre2006,Sayre2007a,Sayre2008,McCann2009}) would have led to different categories. 

Major results of these groupings are shown in tables \ref{technique} and \ref{correctness}. 

\begin{table*}
	\begin{center}
		\begin{tabular}{c|c|c|c|c|c|c|c|c}
		\hline
			& \multicolumn{2}{c}{All years (N=39} \vline & \multicolumn{2}{c}{2006 (N=15)} \vline & \multicolumn{2}{c}{2007 (N=10} \vline & \multicolumn{2}{c}{2008 (N=14} \\
			Category & Midterm & Final & Midterm & Final & Midterm & Final & Midterm & Final \\ 
		\hline
			Used only limits & 24 & 22 & 7 & 8 & 7 & 5 & 10 & 9 \\ 
			Used only +C & 5 & 3 & 2 & 1 & 0 & 0 & 3 & 2 \\ 
			Used limits and +C & 3 & 2 & 3 & 1 & 0 & 1 & 0 & 0 \\ 
			Other & 10 & 12 & 6 & 5 & 3 & 4 & 1 & 3
		\end{tabular}
		\caption{Student solution techniques. Comparing student solution methods (39 matched students) on a question given on a midterm and final}
		\label{technique}
	\end{center}
\end{table*}

\begin{table*}
	\begin{center}
		\begin{tabular}{c|c|c|c|c|c|c|c|c}
		\hline
			& \multicolumn{2}{c}{All years (N=39} \vline & \multicolumn{2}{c}{2006 (N=15)} \vline & \multicolumn{2}{c}{2007 (N=10)} \vline & \multicolumn{2}{c}{2008 (N=14)} \\
			Category & Midterm & Final & Midterm & Final & Midterm & Final & Midterm & Final \\ 
		\hline
			Correct & 6 & 7 & 1 & 2 & 0 & 0 & 5 & 5 \\ 
			Math error & 28 & 18 & 11 & 10 & 10 & 4 & 7 & 4 \\ 
			Bondary condition error & 15 & 16 & 7 & 6 & 3 & 9 & 5 & 1 \\ 
			Other & 4 & 18 & 4 & 6 & 0 & 6 & 0 & 6
		\end{tabular}
		\caption{Student correctness. Comparing student correctness (39 matched students) on a question given on a midterm and final.}
		\label{correctness}
	\end{center}
\end{table*}

\subsection{Apparent conclusions: A short interlude}

Looking at Table \ref{technique}, we see that when all years are taken into consideration (as would typically be done for a course with such low yearly populations and the same instructional techniques), the methods students use remain fairly static, with only a change of one or two students per category. Though we emphasize using the limits method in the course, many students seem to stick to the more familiar +C method that they originally learned in physics. 

Table \ref{correctness} tells a slightly different story, but one that is unfortunately familiar to many instructors: few students moved into the correct category from the midterm to the final exam, but, in contrast, many students' solutions change so much that they must be categorized as ``other." While the number of math errors decreases, the number of boundary conditions errors remains fairly constant. 

An observation might then be:

\begin{quote} 
Instruction did not significantly change the method students use to solve air-resistance problems. While on both midterm and final about 15\% of students are correct, more than twice that many make boundary condition errors, indicative of a disconnect between the mathematical method of separation of variables and the physical reality it represents. Regrettably, students with an unacceptably flawed (categorized as ``other") solution increase from 10\% to 45\%!
 \end{quote}

From these observations, we might pursue several different conclusions. Perhaps the group work preceding the written tests in 2007 and 2008 affected midterm performance and lowered the ``other" responses on the midterm. Perhaps on the final in 2007 and 2008, students simply had too much to study for and were more careless in setting up problems. Other conclusions are available, as well.

Interesting as these hypothetical observations and conclusions might be, more fundamental concerns should be addressed first. The tallies shown in Tables \ref{technique} and \ref{correctness} are incomplete because they cannot tell how the solutions of \textit{individual students} change over the course of the semester. Are those the same five stalwart students getting the right answer on the midterm and final in 2008, or is the old guard getting replaced with a fresh crop? In 2006, fewer students used the ``limits and +C" method on the final, but what happened to those students who abandoned ship? 

A multitude of escalator plots might be created to address these concerns and questions. Answers have serious consequences when considering how to improve teaching. If students who start correct stay correct, we can spend our time focusing on those who need the most help. If students change their answers, then we have to worry about how to help them stay where we wish them to stay. Answers also have serious consequences when considering our research. If students aren't answering consistently, can we fairly use results from a single test or observation? How are we to evaluate students fairly? Some of these concerns are addressed and made clearer when the data are presented in a consistency plot.

\section{\label{Plots}Consistency plots as multi-dimensional analysis}

To show how student responses change over time, we introduce a graphic we call a consistency plot. We draw a \textit{2-d} grid representing two separate ways of analyzing student responses. Each student whose answers are different from midterm to final is shown by an arrow formed of circle, line, and triangle. A circle represents an initial response; a triangle represents a final response. This circle-line-triangle grouping is referred to as a \textit{response pair} and describes the two solution states of a single student at two different times. A square represents a response pair in which the initial and final responses are identical. Both the size of the element and the number placed within it reveal how many students had that particular response pair. 

\begin{figure}[tb]
\begin{center}
\includegraphics[width=3.2in]{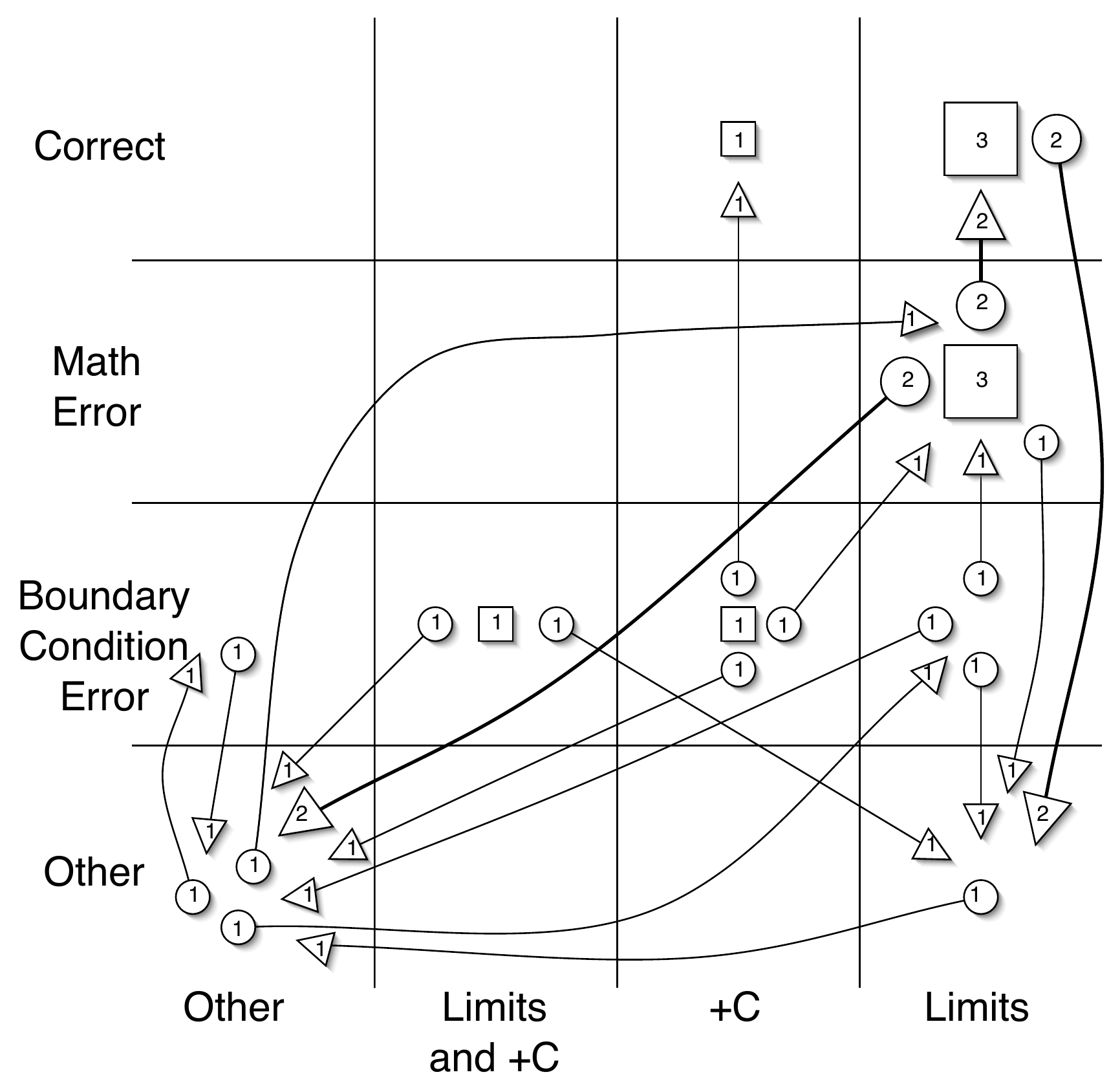}
\caption{Consistency plot showing changes in student midterm and final exam responses to identical questions in both 2006 and 2008.}
\label{Consistency2006-8}
\end{center}
\end{figure}

The plot axes can represent any two lines of analysis of the data; in our case (see figure \ref{Consistency2006-8}), we use the previously discussed categories of solution method and correctness. We have arranged the axes to place those methods we find ``less favorable" closer to what would be the origin in a canonical coordinate system (i.e. the lower left). We place the limits method further to the right because it gives us analytical power that the +C method obscures. For example, when testing the limits of certain constant values in the solution of an integral, the limits method makes the functional form of the equation more apparent sooner than the +C method. But, in situations such as the ``correctness" axis, where categories are not mutually exclusive, we place student responses on the plot according to the most egregious error. For example, a student with both a math error and a boundary condition error appears in the boundary condition error region of the plot. More discussion of this hierarchical coding scheme is found in Section \ref{Tabular}.

Figure \ref{Consistency2006-8} shows a consistency plot of years 2006 and 2008. (2007 was omitted due to the difference in initial and final question statements and will be discussed in Section \ref{Difficulties}) The plot shows a considerable variety in student responses. Most response pairs describe individual students; only five response pairs are made up of more than one student: three students use limits and are correct on both midterm and final; similarly, three use limits and have a math error. Three pairs of students change responses in an identical manner. In total, of the 29 student response pairs in 2006 and 2008, 22 (ca. 3/4) are distinct.  This variability in student responses indicates that there is much more richness than indicated in Tables \ref{technique} and \ref{correctness}. 

\subsection{\label{Elements}Elements of consistency plots}

\begin{figure}[tb]
\begin{center}
\includegraphics[width=3.2in]{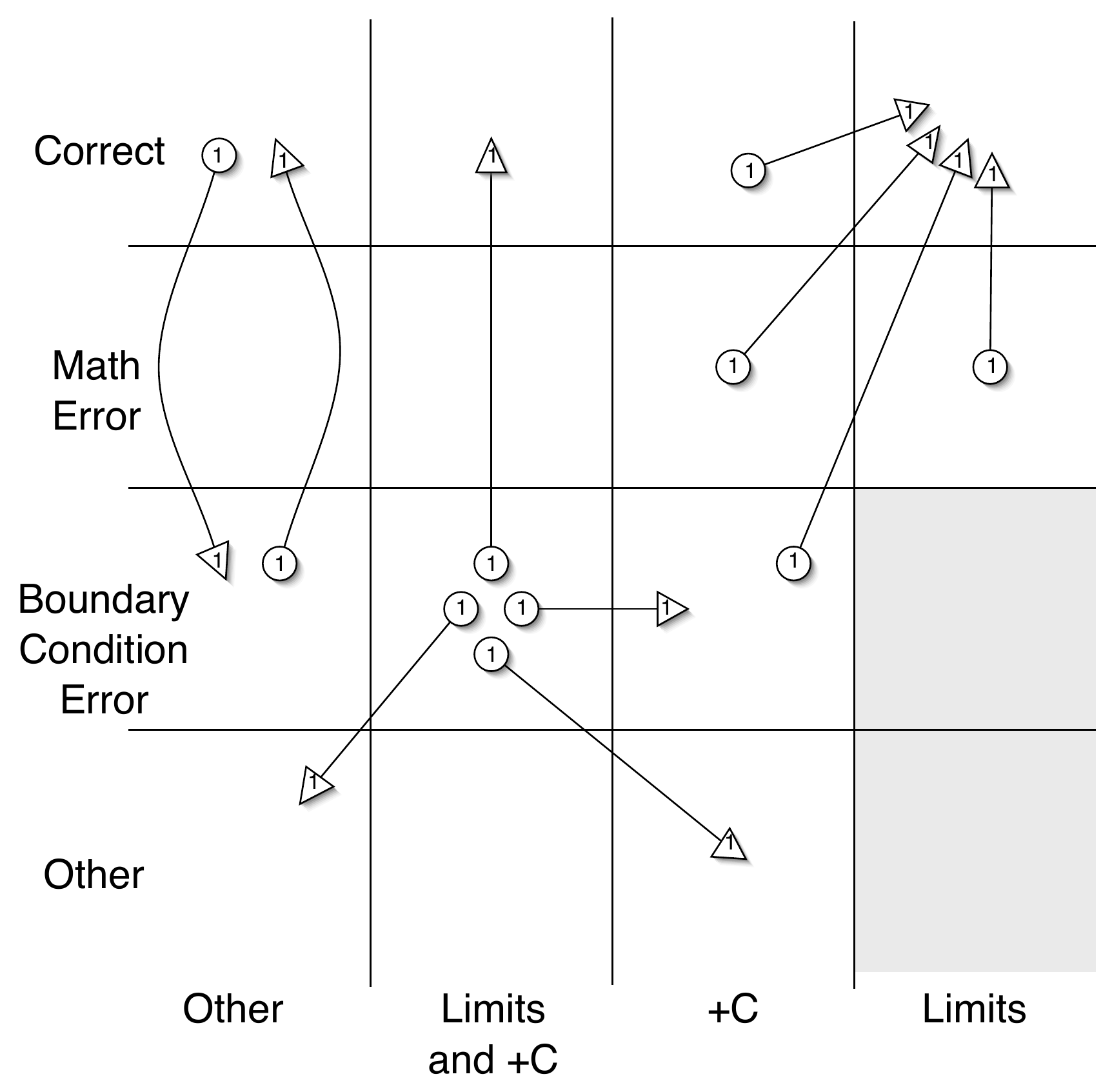}
\caption{Sample consistency plot. Separate examples of voids, starbursts, and attractors are shown.}
\label{ConsistencySample}
\end{center}
\end{figure}

It is in the nature of consistency plots to be at least somewhat messy; the richness of information that they present can look chaotic at first glance. However, patterns do emerge from the plots. Figure \ref{ConsistencySample} shows an idealized consistency plot that contains four elements we have found in our data. 
\begin{itemize}
\item In the upper right corner of the plot, the region (limits, correct) is \textit{attractive}, that is, many more response pairs end there than begin there. 
\item The region (limits and +C, boundary condition error) is an example of a \textit{starburst}, the opposite of an attractor. 
\item To the far left we see \textit{circulation}, where two regions are connected by opposite response pairs. 
\item Finally, in the bottom right, there is a \textit{void}, a region with no responses. 
\end{itemize}

Our idealized consistency plot allowed us to clearly introduce important grouping of student responses, separating each grouping so that they could be clearly identified. As is apparent from our plot of actual data in figure \ref{Consistency2006-8}, real life is not so accommodating. Below, we present examples of these elements from our 2006/08 consistency plot. For clarity, we will show each element present on a separate plot, along with comments regarding the implications of each element for this particular question. Before reading on, however, the reader may find it interesting to return to figure \ref{Consistency2006-8} to find an example of each grouping.

\begin{figure}
\begin{center}
\includegraphics[width=3.2in]{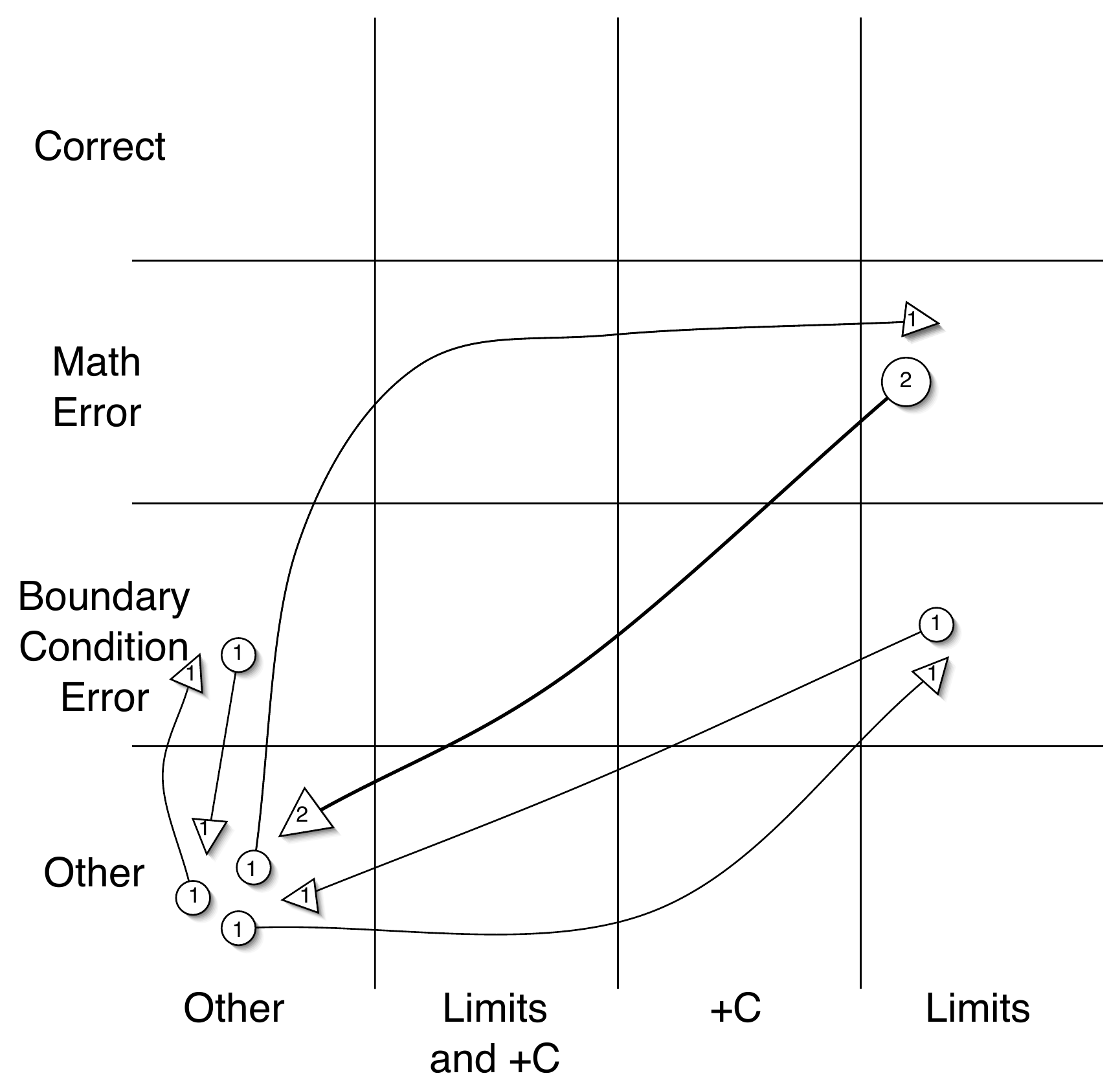}
\caption{Circulation pattern. Students flow out of and into a region on the consistency plot.}
\label{Circulation}
\end{center}
\end{figure}

Figure \ref{Circulation} shows three examples of \textit{circulation}, that is, sets of opposite response pairs. A table-like approach to the data would state that, for example, one student on the midterm and final used an ``other" method while making a boundary condition error. In fact, our consistency plot shows that this data point is made up of two students, each of whom at one time or another was also in the (other, other) category. Similarly, while a table would show a net flow (of one for the data presented on this sub-plot) into the ``other, other" category it would not show that three students improved their responses. Circulation is the 2-D version of Kanim's escalator diagram. Examples of circulation of are of interest since these are the changes in response that are absolutely undetectable in a traditional reporting of data, and are often ignored when matched student responses are not considered.

\begin{figure}
\begin{center}
\includegraphics[width=3.2in]{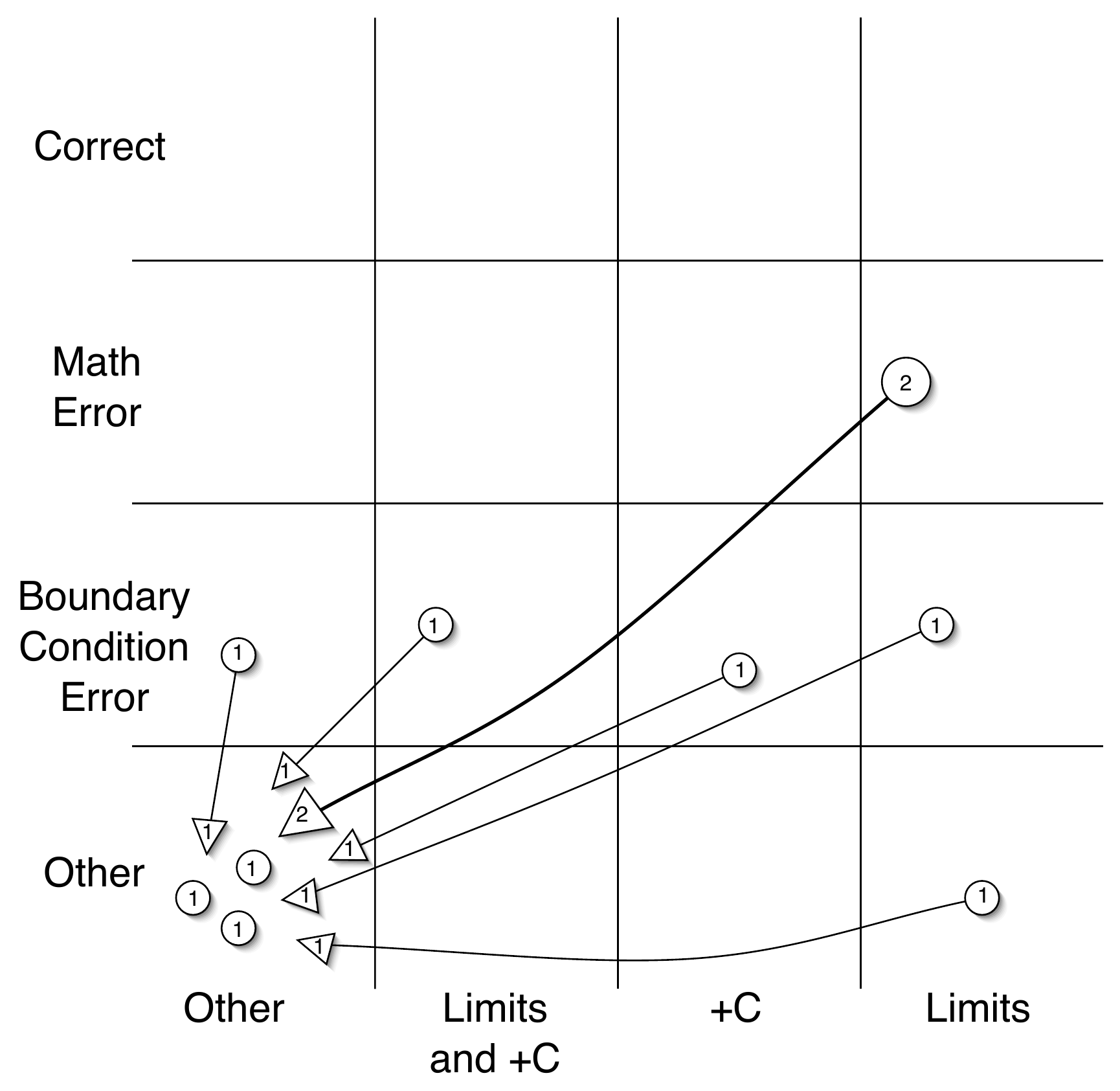}
\caption{Attractor pattern. Students move toward one region on the consistency plot.}
\label{Attractor}
\end{center}
\end{figure}

In reality, there was much more flow into the (other, other) region than out (see figure \ref{Attractor}). We call such a region an \textit{attractor}. In our idealized plot, no responses left the region; in this real world example, three leave, giving a net flow of four students. Seven students found themselves in this category on the final. Yet, all but one of the midterm solutions used a separation of variables technique, and three contained only math errors. In other words, most of the students who performed badly on the final gave little forecast of this eventuality on the midterm. 

Although we would much rather see the (limits, correct) region be the attractor, (as it was on our idealized plot, figure \ref{ConsistencySample}) this attractor makes visual the unfortunate case of students who seemed to have it together during the course falling apart on a final examination that covers many topics. These results call into question the ways in which we use individual (non-linked) midterm and final examinations to make claims about what students do and do not know.

\begin{figure}
\begin{center}
\includegraphics[ width=3.2in]{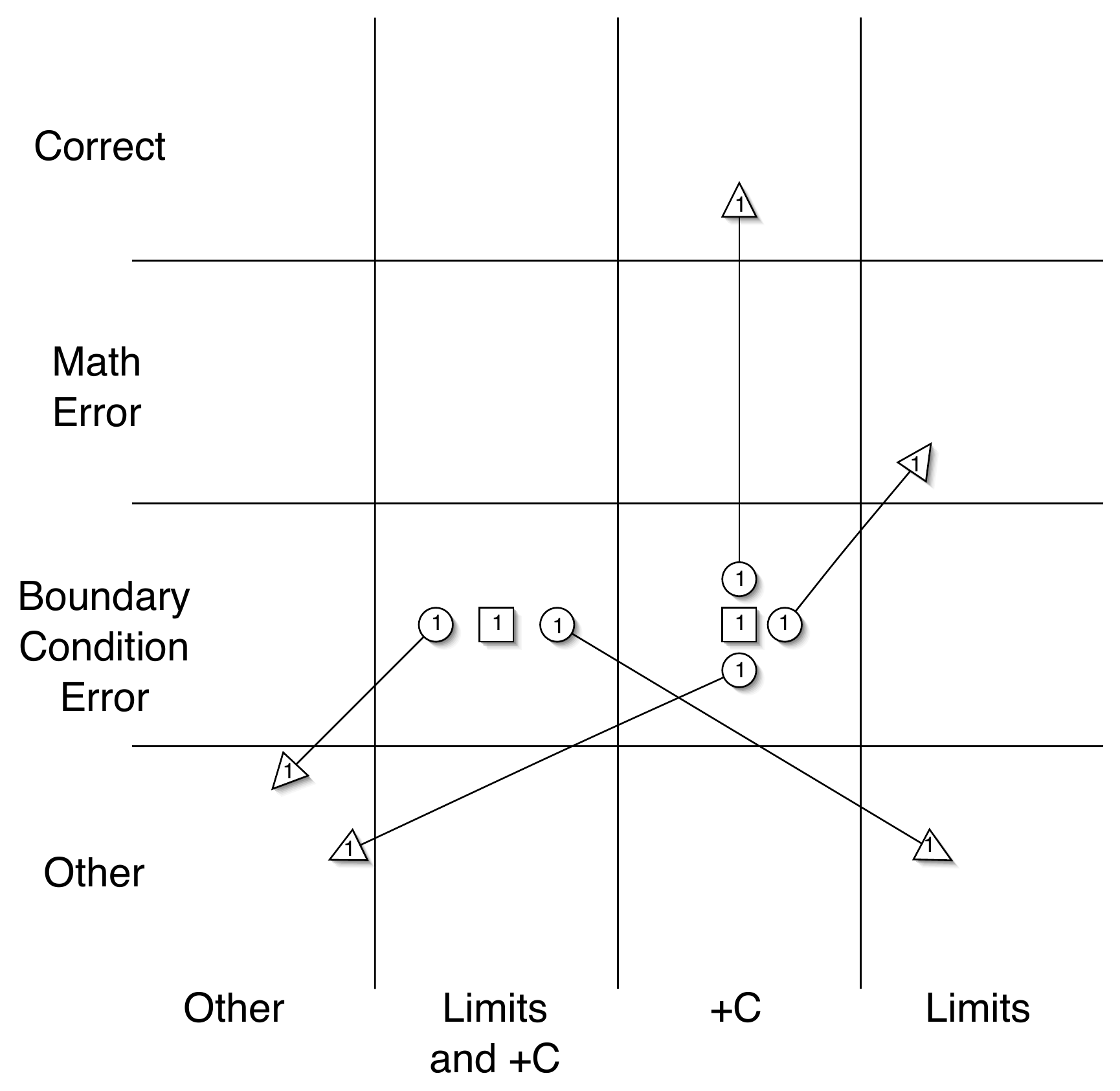}
\caption{Starburst pattern. (Nearly all) students leave a given region on the consistency plot.}
\label{Starbursts}
\end{center}
\end{figure}

Two small starbursts appear on our 2006/08 consistency plot (see figure \ref{Starbursts}), with origins in (+C, boundary condition error) and (limits and +C, boundary condition error). In each case, two important aspects are apparent: no students enter the region, and the students that leave go in all directions, rather than to another specific region. This implies that these combinations of methods and errors are the result of unrefined (``everything plus the kitchen sink") thinking about the problem. Had the students all gone in the same direction, we might instead infer that the indicated region can be thought of as a stepping-stone toward a more refined solution. 

We can see the difference between the ``kitchen-sink" and the ``stepping stone" interpretation when we look at the starburst in the (limits and +C, boundary condition error) region. Three students use both limits and an integration constant in their solutions on the midterm with two changing their methods on the final. 
When we first ran across this method on the midterm, we interpreted it as a bridge or stepping stone between the use of integration constants, the mainstay of math classes, and integration limits, modeled by the physics instructor. Although mathematically sketchy (since, of course, limits and an integration constant serve the same mathematical role), we thought that it was simply a sign of students becoming used to the limits method and including it in their solutions without yet recognizing that the two methods were mathematically equivalent. We suspected that if this were the case, these students would later solely use the limits method as they became more comfortable with it. Had our assumptions been correct, these three students would have all moved toward the (limits, correct) region. They did not. The starburst indicates, instead, that these students were, on the midterm, simply throwing all the techniques available to them at the problem, and did the same on the final. The one student who does switch to the limits method makes errors substantial enough to be considered ``other."
 
\begin{figure}
\begin{center}
\includegraphics[width=3.2in]{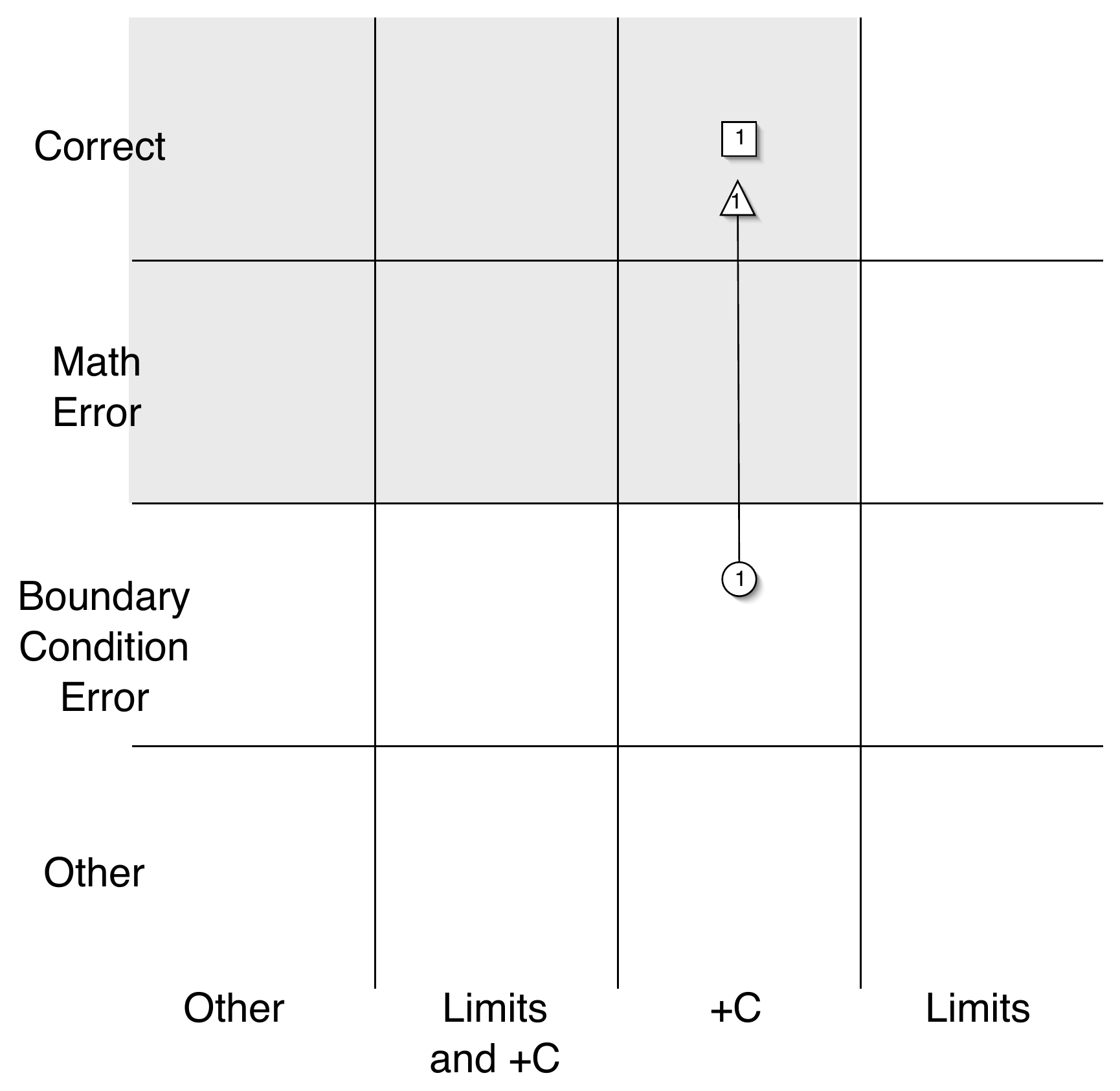}
\caption{Void pattern. (Nearly) no answers are found in an extended region of the consistency plot.}
\label{Void}
\end{center}
\end{figure}

Finally, we consider in figure \ref{Void} the highlighted \textit{void} region where few student responses are located. There are 21 solutions in the region below the void, representing students who used a method other than limits and either did not successfully incorporate the boundary condition into the problem or were unable to appropriately represent the physical situation mathematically. Of 29 students over two exams in two years, only two (a total of three times) use an integration constant and find its value using the boundary condition. No students correctly used a method other than separation of variables, though other appropriate techniques exist. One might recall the form of the solution of the differential equation or use the more complex technique of variation of parameters. Of course, zeros show up in tables, but the visual nature of the consistency plot makes the regions that contain few students as compelling as the regions that contain many students, and leads us to consider what students are not doing along with what they are doing. Since we are looking at how students bring physical meaning to the mathematics, the fact that only two of 29 students use a method other than the limits method to correctly solve the problem is noteworthy and suggests several pedagogical pathways toward helping students learn the physics (and mathematics) better.

\subsection{\label{Difficulties}Difficulties with a New Representation}

As we have shown, our representation can allow a deeper analysis of data than a simple table. However, it is not without difficulties. As an example, we present a consistency plot of student responses during the year 2007. In this year, the problem was changed on the final so that the positive direction was up (in figure \ref{Problem} it was down). Our goal in reversing the direction was to give a subtly different problem from the midterm, so that memorized responses couldn't be used as readily. The change in coordinate system leads to a change in the definition of the initial velocity of the downwardly thrown ball, among other things. Since, by convention, physical constants are positive, students needed to define the initial velocity of the ball as $-v_0$. The change also caused serious problems for students not setting up the problem correctly; their translation of the physical system into coordinates often included the wrong signs. Such errors were considered ``other" errors (in keeping with our previous definitions). Because of the seriousness of the error, students making ``other" errors would not be counted as making ``boundary condition" or ``math" errors. 

\begin{figure}
\begin{center}
\includegraphics[width=3.2in]{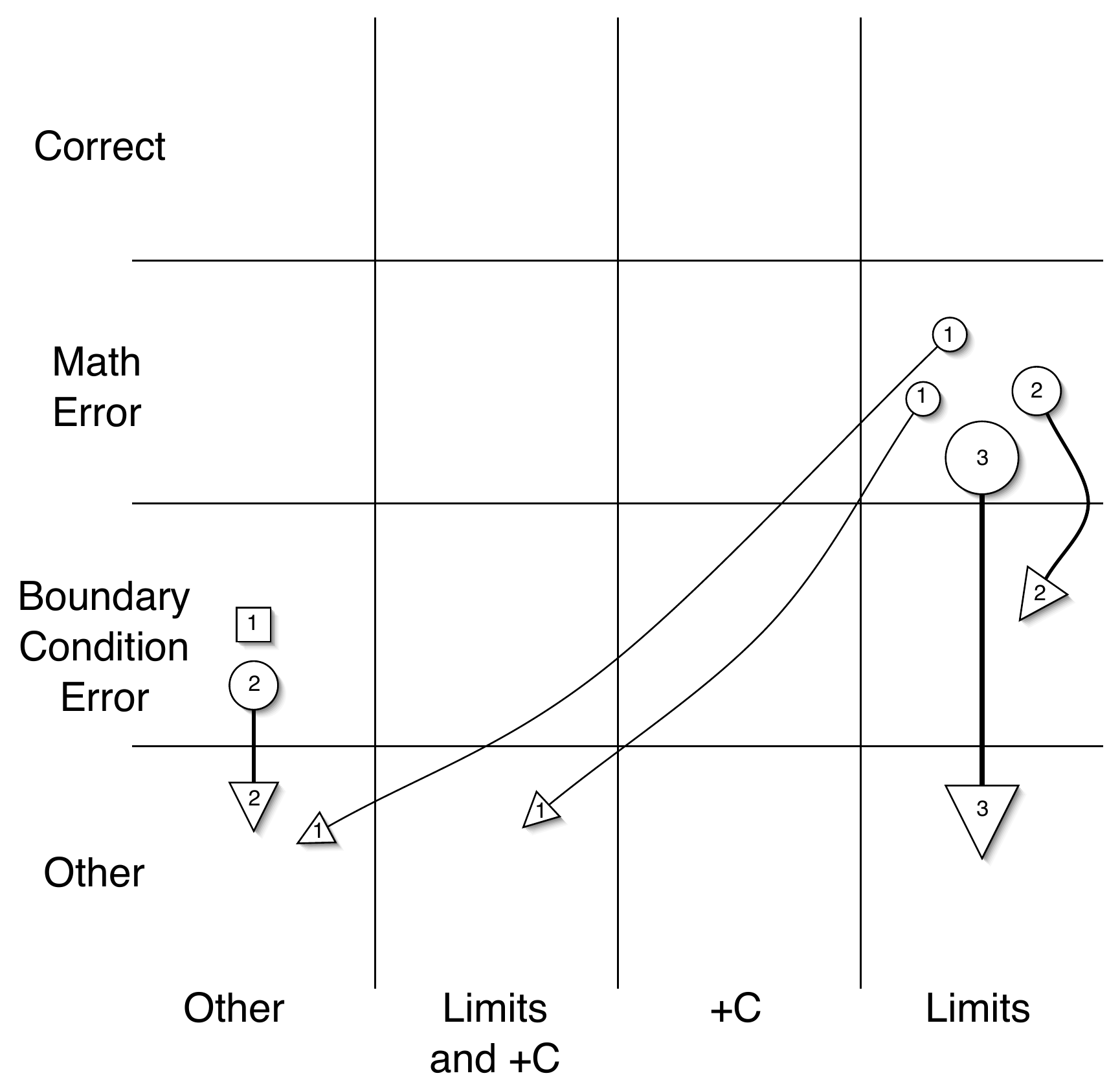}
\caption{Consistency plot from 2007. Final examination question had a differen coordinate systems, leading to many ``other" responses that hide information from further analysis.}
\label{Consistency2007}
\end{center}
\end{figure}

An initial perusal of figure \ref{Consistency2007} suggests that only three students ended in the ``boundary condition error" row of the diagram. However, the hierarchical nature of our vertical axis obscures the fact presented in table \ref{correctness} that nine students actually made the $-v_0$ error. Because seven students made errors in the setup and execution of the problem that were severe enough to classify as ``other," we lose information about the prevalence of boundary condition errors in the 2007 consistency plot.

In part, this situation can be alleviated by careful attention to coding and the development of coding schemes. However, we cannot always predict in advance what the data might say and there is always the potential that a hierarchical coding scheme will conceal interesting patterns, so we do not recommend the abandonment of the data table. One might allow for fuzzy categories, but having one student in two regions seems impossible to interpret visually. One might create more detailed exclusive categories, but developing too many categories to eliminate a hierarchical axis and describe student responses more fully may create a plot too chaotic to read. We already have very few students doing the same thing from midterm to final -- to use a finer comb to separate the responses would make the data (including observation of attractors and starbusts) nearly meaningless for small classes. The balance between inclusiveness of categories and visual accessibility will be specific to each data set, and so must be determined on an individual basis. We have chosen to keep the 2007 consistency plot on the same scale as the 2006/08 plot, for example.

\section{Conclusions}

We have described consistency plots, a new method of presenting student responses and how they change over time. As a visual presentation of data, they allow the recognition of patterns of student response that may not be available or easily discernable when the data is presented in table form. As an example, we use the plots to discuss how students' solution methods and correctness for solving first-order separable differential equations change from a midterm to a final exam. 

We show several examples of interesting response patterns within the overall plot: circulation, attractors, starbursts, and voids. Each is revealing about student performance in the class, yet most are hard to represent in simple tabular form. Circulation calls into question how one uses class-wide tables of data to imply improvement in student performance. Attractors can show where students are being led over the course of instruction. Starbursts help us distinguish between solutions that indicate students developing toward new ideas (i.e., stepping stones) and solutions that indicate students' confusion about the problem. Voids indicate a ``failure" to use some method correctly, but give little further information. 

We also note difficulties with the use of consistency plots. When the hierarchical categorization of data is employed, some patterns of student response may be obscured. The inability to use fuzzy categorization and requiring students to be in only one region is another problem. Student responses are often richer than a simple plot can represent. Thus, consistency plots are limited by the researchers' assumptions and interpretations of the data. Our interests in this paper were in students giving physical meaning to mathematical solutions. Different plots would have been drawn had we been analyzing how students translated a physical situation into mathematical statements.
	
We believe consistency plots can be useful for displaying data in a wide variety of situations. Whenever individual student responses to pre- and post test questions are being considered, it is likely that a consistency plots will be a useful addition to standard tables for interpreting patterns of student response.

\appendix


\bibliographystyle{apsrev-stper}
\bibliography{MichaelBibDesk}

\begin{thebibliography}{14}
\expandafter\ifx\csname natexlab\endcsname\relax\def\natexlab#1{#1}\fi
\expandafter\ifx\csname bibnamefont\endcsname\relax
  \def\bibnamefont#1{#1}\fi
\expandafter\ifx\csname bibfnamefont\endcsname\relax
  \def\bibfnamefont#1{#1}\fi
\expandafter\ifx\csname citenamefont\endcsname\relax
  \def\citenamefont#1{#1}\fi
\expandafter\ifx\csname url\endcsname\relax
  \def\url#1{\texttt{#1}}\fi
\expandafter\ifx\csname urlprefix\endcsname\relax\def\urlprefix{URL }\fi
\providecommand{\bibinfo}[2]{#2}
\providecommand{\eprint}[2][]{\url{#2}}

\bibitem[{\citenamefont{Ambrose}(2004)}]{Ambrose2004}
\bibinfo{author}{\bibfnamefont{B.~S.} \bibnamefont{Ambrose}},
  \emph{\bibinfo{title}{Investigating student understanding in intermediate
  mechanics: Identifying the need for a tutorial approach to instruction}},
  \bibinfo{journal}{American Journal of Physics} \textbf{\bibinfo{volume}{72}},
  \bibinfo{pages}{453} (\bibinfo{year}{2004}).

\bibitem[{\citenamefont{Wittmann and Ambrose}()}]{Wittmann2007b}
\bibinfo{author}{\bibfnamefont{M.~C.} \bibnamefont{Wittmann}} \bibnamefont{and}
  \bibinfo{author}{\bibfnamefont{B.~S.} \bibnamefont{Ambrose}},
  \emph{\bibinfo{title}{Intermediate {M}echanics {T}utorials:
  http://perlnet.umaine.edu/imt/}},
  \urlprefix\url{http://perlnet.umaine.edu/imt/}.

\bibitem[{\citenamefont{Hestenes et~al.}(1992)\citenamefont{Hestenes, Wells,
  and Swackhamer}}]{Hestenes1992}
\bibinfo{author}{\bibfnamefont{D.}~\bibnamefont{Hestenes}},
  \bibinfo{author}{\bibfnamefont{M.}~\bibnamefont{Wells}}, \bibnamefont{and}
  \bibinfo{author}{\bibfnamefont{G.}~\bibnamefont{Swackhamer}},
  \emph{\bibinfo{title}{Force concept inventory}}, \bibinfo{journal}{The
  Physics Teacher} \textbf{\bibinfo{volume}{30}}, \bibinfo{pages}{141}
  (\bibinfo{year}{1992}).

\bibitem[{\citenamefont{Thornton and Sokoloff}(1998)}]{Thornton1998}
\bibinfo{author}{\bibfnamefont{R.~K.} \bibnamefont{Thornton}} \bibnamefont{and}
  \bibinfo{author}{\bibfnamefont{D.~R.} \bibnamefont{Sokoloff}},
  \emph{\bibinfo{title}{Assessing student learning of Newton's laws: The Force
  and Motion Conceptual Evaluation and the Evaluation of Active Learning
  Laboratory and Lecture Curricula}}, \bibinfo{journal}{American Journal of
  Physics} \textbf{\bibinfo{volume}{66}}, \bibinfo{pages}{338}
  (\bibinfo{year}{1998}).

\bibitem[{\citenamefont{McDermott and Redish}(1999)}]{McDermott1999}
\bibinfo{author}{\bibfnamefont{L.~C.} \bibnamefont{McDermott}}
  \bibnamefont{and} \bibinfo{author}{\bibfnamefont{E.~F.}
  \bibnamefont{Redish}}, \emph{\bibinfo{title}{Resource Letter PER-1: Physics
  Education Research}}, \bibinfo{journal}{American Journal of Physics}
  \textbf{\bibinfo{volume}{67}}, \bibinfo{pages}{755} (\bibinfo{year}{1999}),
  \urlprefix\url{http://www.physics.umd.edu/perg/papers/redish/rl.pdf}.

\bibitem[{\citenamefont{Hake}(1998)}]{Hake1998}
\bibinfo{author}{\bibfnamefont{R.~R.} \bibnamefont{Hake}},
  \emph{\bibinfo{title}{Interactive-engagement versus traditional methods: A
  six-thousand-student survey of mechanics test data for introductory physics
  courses}}, \bibinfo{journal}{American Journal of Physics}
  \textbf{\bibinfo{volume}{66}}, \bibinfo{pages}{64} (\bibinfo{year}{1998}).

\bibitem[{\citenamefont{Redish}(1999)}]{Redish1999}
\bibinfo{author}{\bibfnamefont{E.~F.} \bibnamefont{Redish}},
  \emph{\bibinfo{title}{Millikan Award Lecture (1998): Building a Science of
  Teaching Physics}}, \bibinfo{journal}{American Journal of Physics}
  \textbf{\bibinfo{volume}{67}}, \bibinfo{pages}{562} (\bibinfo{year}{1999}).

\bibitem[{\citenamefont{Bao}(2006)}]{Bao2006a}
\bibinfo{author}{\bibfnamefont{L.}~\bibnamefont{Bao}},
  \emph{\bibinfo{title}{Theoretical Comparison of Average Normalized Gain
  Calculations}}, \bibinfo{journal}{American Journal of Physics}
  \textbf{\bibinfo{volume}{74}}, \bibinfo{pages}{917} (\bibinfo{year}{2006}).

\bibitem[{\citenamefont{Bao and Redish}(2006)}]{Bao2006}
\bibinfo{author}{\bibfnamefont{L.}~\bibnamefont{Bao}} \bibnamefont{and}
  \bibinfo{author}{\bibfnamefont{E.~F.} \bibnamefont{Redish}},
  \emph{\bibinfo{title}{Model analysis: Representing and assessing the dynamics
  of student learning}}, \bibinfo{journal}{Physical Review Special Topics
  Physics Education Research} \textbf{\bibinfo{volume}{2}},
  \bibinfo{pages}{010103} (\bibinfo{year}{2006}).

\bibitem[{\citenamefont{{Van Deventer}}(2008)}]{VanDeventer2008}
\bibinfo{author}{\bibfnamefont{J.}~\bibnamefont{{Van Deventer}}},
  \bibinfo{type}{unpublished {M}aster of {S}cience in {T}eaching thesis},
  \bibinfo{school}{University of Maine}, \bibinfo{address}{Orono, ME}
  (\bibinfo{year}{2008}).

\bibitem[{\citenamefont{McCann and Wittmann}(2009)}]{McCann2009}
\bibinfo{author}{\bibfnamefont{K.}~\bibnamefont{McCann}} \bibnamefont{and}
  \bibinfo{author}{\bibfnamefont{M.~C.} \bibnamefont{Wittmann}},
  \emph{\bibinfo{title}{The role of sign in students' modeling of scalar
  equations}}, \bibinfo{journal}{The Physics Teacher}
  \textbf{\bibinfo{volume}{accepted for publication}} (\bibinfo{year}{2009}).

\bibitem[{\citenamefont{Sayre et~al.}(2007)\citenamefont{Sayre, Wittmann, and
  Donovan}}]{Sayre2006}
\bibinfo{author}{\bibfnamefont{E.~C.} \bibnamefont{Sayre}},
  \bibinfo{author}{\bibfnamefont{M.~C.} \bibnamefont{Wittmann}},
  \bibnamefont{and} \bibinfo{author}{\bibfnamefont{J.~E.}
  \bibnamefont{Donovan}}, \emph{\bibinfo{title}{Resource Plasticity: Detailing
  a Common Chain of Reasoning with Damped Harmonic Motion}}, in
  \emph{\bibinfo{booktitle}{Physics Education Research Conference 2007, AIP
  Conference Proceedings 883}}, edited by
  \bibinfo{editor}{\bibfnamefont{L.}~\bibnamefont{McCullough}},
  \bibinfo{editor}{\bibfnamefont{L.}~\bibnamefont{Hsu}}, \bibnamefont{and}
  \bibinfo{editor}{\bibfnamefont{P.~R.} \bibnamefont{Heron}}
  (\bibinfo{publisher}{Springer New York, LLC}, \bibinfo{address}{Secaucus,
  NJ}, \bibinfo{year}{2007}), vol. \bibinfo{volume}{883}, pp.
  \bibinfo{pages}{85--88}.

\bibitem[{\citenamefont{Sayre and Wittmann}(2007)}]{Sayre2007a}
\bibinfo{author}{\bibfnamefont{E.~C.} \bibnamefont{Sayre}} \bibnamefont{and}
  \bibinfo{author}{\bibfnamefont{M.~C.} \bibnamefont{Wittmann}},
  \emph{\bibinfo{title}{Intermediate mechanics students' coordinate system
  choice}}, in \emph{\bibinfo{booktitle}{Conference on Research in
  Undergraduate Mathematics Education}}, edited by
  \bibinfo{editor}{\bibfnamefont{M.}~\bibnamefont{Oehrtman}}
  (\bibinfo{publisher}{SIGMAA}, \bibinfo{year}{2007}).

\bibitem[{\citenamefont{Sayre and Wittmann}(2008)}]{Sayre2008}
\bibinfo{author}{\bibfnamefont{E.~C.} \bibnamefont{Sayre}} \bibnamefont{and}
  \bibinfo{author}{\bibfnamefont{M.~C.} \bibnamefont{Wittmann}},
  \emph{\bibinfo{title}{Plasticity of intermediate mechanics students\char39{}
  coordinate system choice}}, \bibinfo{journal}{Phys. Rev. ST Phys. Educ. Res.}
  \textbf{\bibinfo{volume}{4}}, \bibinfo{pages}{020105} (\bibinfo{year}{2008}).

\end{thebibliography}

\end{document}